\documentclass{iau}

\usepackage{amsmath}
\usepackage{graphicx}
\usepackage{multirow}

\begin{document}

\lefttitle{Di Vruno et al.}
\righttitle{Modelling satcons radio emission}

\jnlPage{1}{7}
\jnlDoiYr{2024}
\doival{10.1017/xxxxx}

\aopheadtitle{Proceedings IAU Symposium }
\editors{eds.}

\title{Modelling steerable beams of satellite constellations for radio astronomy impact studies}

\author{Federico~Di~Vruno$^{1,2}$, Gary~Hovey$^{3}$}
\affiliation{1. SKA Observatory, Jodrell Bank, Macclesfield SK11 9FT, United Kingdom. \email{federico.divruno@skao.int}}
\affiliation{2. IAU Centre for the Protection of the Dark and Quiet Sky from Satellite Constellation Interference, France.}
\affiliation{3. Onsala Space Observatory, Chalmers University of Technology  412 96 Gothenburg , Sweden.}

\begin{abstract}
Modelling is essential for studies that quantify the impact from satellite downlinks on radio astronomy sites. To estimate this impact it is necessary to know not only the position and velocity of satellites but also their behaviour in the radio spectrum domain. As many large satellite constellations are using steerable beam antennas, deterministically predicting the transmitted power towards a defined direction (in this case where a radio telescope points) becomes an almost impossible task and therefore another approach has to be used. This work presents a method to simulate and estimate the percentiles of the radiation pattern of satellites with steerable beam patterns based on simulations and a comparison with measurements of Starlink satellites using the Onsala Twin Telescopes in Sweden.
\end{abstract}


\maketitle

\section{Introduction}
Large constellations of satellites deployed in Low Earth Orbit (LEO) have been a concern due to their impact on optical and radio astronomy and the night sky as discussed in \citep{walker_impact_2020}, \citep{hall_satcon2_2021},  \citep{iau_dark_2021}, \citep{lawrence_case_2022}, and \citep{walker_dark_2022} . For the case of the impact on radio astronomy, few studies have been published, like \citep{ecc_report_271}, \citep{di_vruno_large_2023} or in \citep{Jasons_report_satellite_constellations}. These studies considered the impact of the largest constellations being deployed, such as Starlink and OneWeb, but they have either not considered the strong in-band emissions from satellites or used significant simplifications when referring to satellite's emissions. 

None of the reports mentioned have considered the complexity of a satellite system such as the Starlink, which uses steerable beams with footprints of 24 km diameter on the ground when considering the -3dB power flux density level \footnote{\url{https://fcc.report/IBFS/SAT-MOD-20200417-00037/2274316}}. While an observer on the ground can estimate the position of most satellites in the sky within a certain accuracy, there is no simple way of knowing how many beams per channel is a particular satellite producing and where are those beams are pointing to at a specific time.

This paper presents the results of a model used to simulate the statistics of the power flux density (PFD) produced by a Starlink satellite and results of measurements conducted with the Onsala Twin Telescopes at the Onsala Space Observatory in Sweden.

\section{Motivation}
Radio telescopes rely on a clean radio spectrum to observe the faintest radio sources of cosmic origin. Within the radio frequency spectrum (from 30 kHz up to 3000 GHz according to the International Telecommunication Union) there are several frequency bands that are allocated to the radio astronomy service (RAS). These "reserved" frequency bands are spread throughout the radio spectrum and centred either in frequencies where there are important astronomical emissions, like the Hydrogen line in 1420.4057 MHz with has allocated frequencies between 1400 to 1427 MHz, or allocated roughly every frequency octave to allow an efficient sampling of a very broadband continuum emission. These radio astronomy bands were defined back in the 1970s and have been fundamental for the RAS since then, but with the advancement of technology and scientific techniques, there is an ever pressing need to use wider frequency bands and increase the sensitivity of instruments. This situation of course collides with the reality that the frequency bands not allocated to radio astronomy are heavily populated with radio signals from other users of the radio spectrum such as cell phones, WIFI and Bluetooth, and satellite transmissions. When any of these man-made signals is received by extremely broadband radio astronomy receivers it's flagged as "interference" by astronomers, while it is understandable that from the point of view of an astronomer these are interfering signals, it is necessary to recognise that often these are legitimate transmissions within the allocated frequency bands for those services.

One effective way of widening the available radio spectrum for radio astronomy is to locate radio telescopes in extremely remote areas, where the terrestrial use of the radio spectrum is minimal. This is the case for telescopes such as the Square Kilometre Observatory (see \citep{highlights_SKA_Mid} and \citep{highlights_SKA_Low}) which consists of two radio telescopes, one in South Africa and the other in Western Australia, both located in nationally protected Radio Quiet Zones (RQZ). An RQZ is a nationally defined area where the use of the radio spectrum, in a very large swath of frequencies, is especially managed to benefit radio astronomy observations (see \citep{itu_ra2259_1}). One of the limitations of RQZs is that their regulations are not effective against satellite transmissions. The recent development of large satellite constellations like Starlink and OneWeb has risen concerns in the radio astronomy community as telescopes can be impacted not only by the "in-band" downlinks of the satellites (their rightfully allocated frequencies), but by low-level "out-of-band" emissions (meaning emissions outside of their allocated frequencies as a sub product of the in-band transmissions), and by unintended electromagnetic radiation (or UEMR) \citep{di_vruno_unintended_2023}.

\begin{figure}[h]
  \center
  \includegraphics[width=0.75\linewidth]{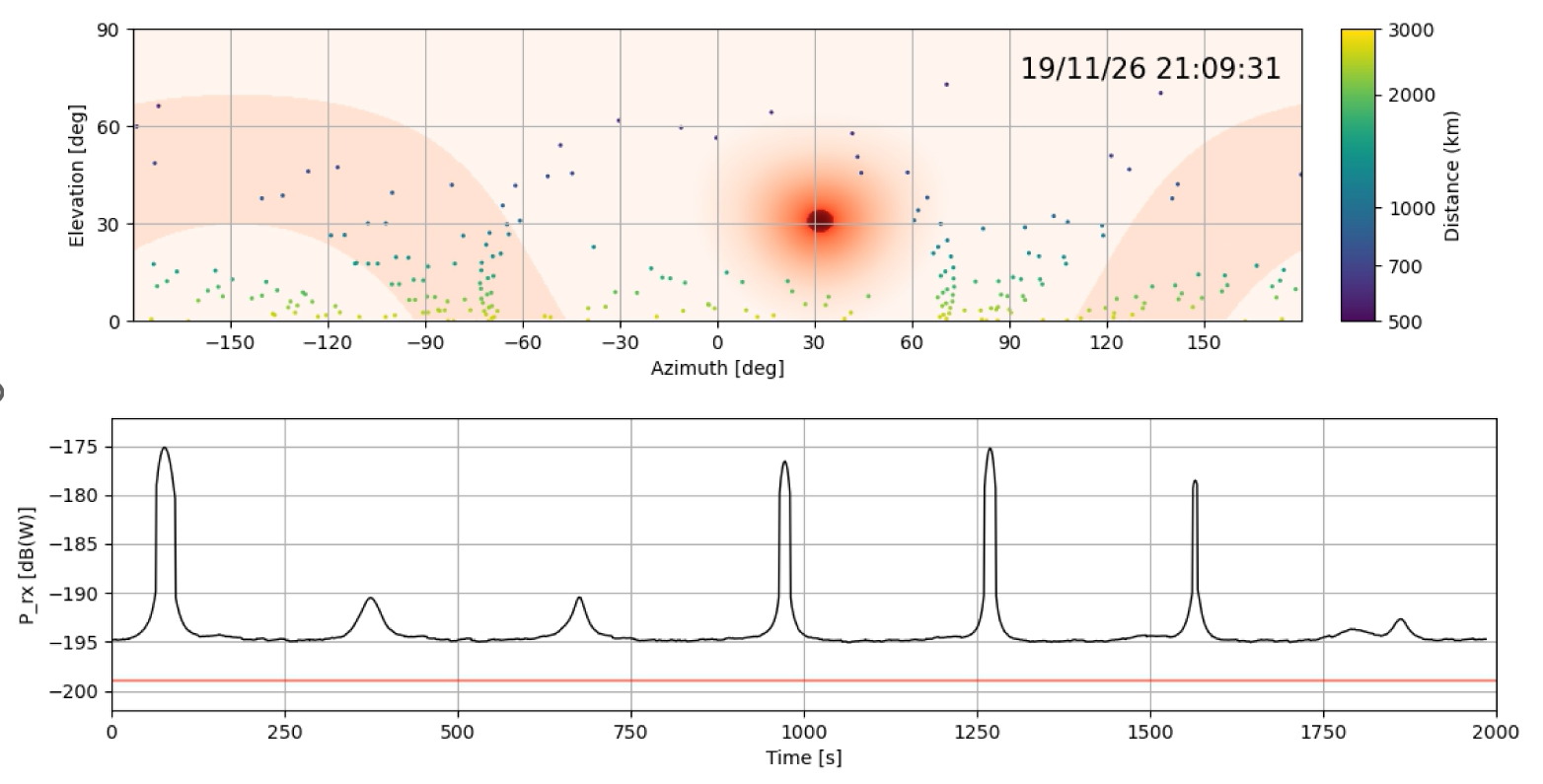}
  \caption{Instantaneous view of satellites in the sky (in azimuth and elevation) color coded in range to the observer, the shaded areas depict the gain pattern defined in recommendation ITU-R RA.1610 (top). Time domain total power received by the radio telescope, the high peaks are when a satellite transits the boresight (bottom). Source: Benjamin Winkel, MPIfR - CRAF }
  \label{fig:epfd}
\end{figure}

The impact of large satellite constellations (on nonGSO constellations as defined by the ITU) on radio telescopes is calculated by using the equivalent power flux density (EPFD) method, this is a computational method that considers the movement of satellites in the local sky, their radio frequency emissions, the reception characteristics of the observer (in this case a radio telescope) and yields the instantaneous and integrated received power (see Fig.~\ref{fig:epfd}). The EPFD method was developed by the ITU to address the situation that all satellites in a nonGSO constellation are constantly moving in the sky, therefore the impact into an observer must be calculated taking this into account. Radio telescope observations differ significantly from a typical ground terminal of a satellite constellation, especially in their sensitivity, variable pointing directions and integration times, therefore the EPFD method to consider the impact on  radio astronomy from nonGSO satellites is considered in \citep{ITU_R_rec_s_1586} and \citep{ITU_R_rec_m_1583}. A detailed description of the EPFD method for radio astronomy can also be found in section 3 [check] of \citep{di_vruno_unintended_2023}. 

The EPFD of a satellite constellation is a function of the following:
\begin{itemize}
    \item The position of the observer in latitude, longitude, and height
    \item The complete radiation pattern of the radio telescope (or using a model)
    \item The position of all satellites in the constellation as a function of time
    \item The radiation characteristics of each satellite (centre frequency, bandwidth, radiation pattern characteristics, etc)
\end{itemize}

Given these parameters the EPFD distribution, in \(\frac{W}{m^2.Hz}\),  may be computed for each cell above the local horizon. With this distribution the CDF and other statistics maybe computed and compared with thresholds for harmful interference (such as defined in \citep{itu_ra769_2}).  

\subsection{Spectral flux density and steerable beams}
\begin{figure}[h]
  \center
  \includegraphics[width=0.5\linewidth]{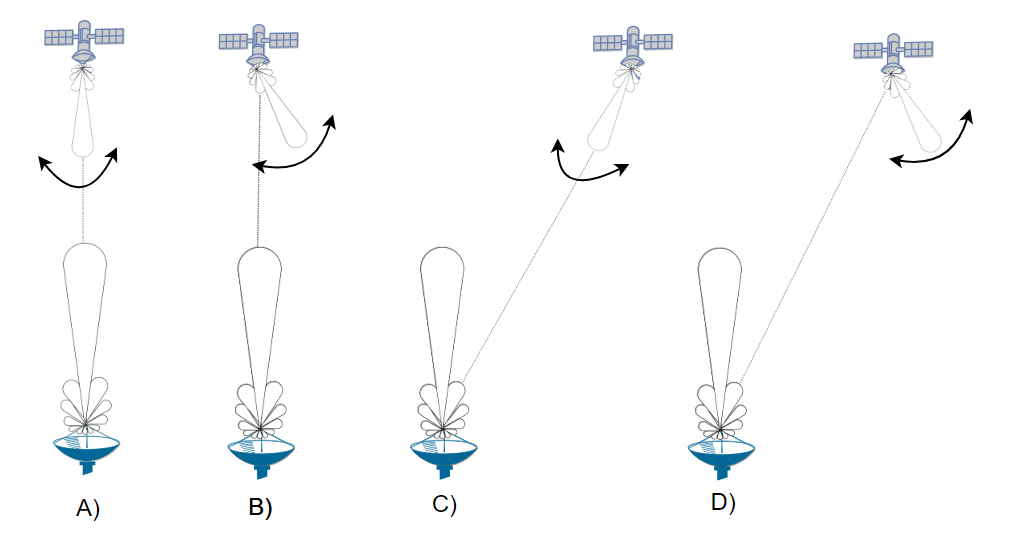}
  \caption{Radio telescope and satellite pointing possible scenarios when the satellite has steerable beams}
  \label{fig:pointing_scenarios}
\end{figure}

While the first three parameters required to calculate the EPFD are readily available, the radiation characteristics of satellites with steerable beams, such as Starlink or Amazon Kuiper, have a radiation patterns that is not static in time. Similarly their spectral usage is also dynamic as their downlink transmission are spread spectrum over the range of 10.7 to 12.7 GHz, divided into eight 250 MHz frequency channels. By using electronically phased transmitters these satellites can produce multiple beams in each frequency channel with relatively small footprints (24 km diameter in the case of Starlink) as well as changing their pointing direction in short timescales to service different cells on the ground. As depicted in Fig. \ref{fig:pointing_scenarios} radio telescope and the satellite constellations interact in four different ways making it impossible for the astronomer to accurately know the pointing direction of each antenna beam and therefore its impact on the observation. Note that this analysis is valid for in-band transmissions, where the beam-forming is designed to operate. The out-of band response could be quite different and futher complicates the situation as the information is not available. In the case of UEMR, the emission is not even produced by the communication antenna so beam forming is out of the question.

The power received by the radio telescope from the satellite is maximum on-axis when both beams are pointing at each other and is reduced as either is pointed away from the other as shown in Fig \ref{fig:pointing_scenarios}. The probability of a satellite being on-axis with a radio telescope is proportional to the number of satellites in LEO, their orbital characteristics, and the longitude of the observer. This probability continues to increase as more satellites are deployed. 

It is important to understand that the purpose of LEO satellite's downlink signals is to communicate reliably (with high signal to noise ratio, SNR)  from space to earth stations some 500 km away. This requires high power levels which are constraint by the maximum power flux density (PFD) allowed by regulations. The maximum power flux density (PFD) that Starlink has declared in their 2020 FCC filling is  \(-146~dB\frac{W}{m^2.4kHz}\), produced at the boresight of the steerable beam or in the "cell" that the beam is illuminating. This is equivalent to  \(-182~dB\frac{W}{m^2.Hz}\) , considering that emissions are uniform in a 4kHz band, and is millions of times stronger than a typically strong radio astronomy source at these frequencies. Consequently, LEO satellites can easily drive a radio-telescope signal chain into non-linear operation if the two beams are within a few beamwidths of each other, because even radio telescope antennas of a modest diameter has a gain of 60 dB or more at these frequencies.

\section{Model}
\begin{figure}[h]
  \center
  \includegraphics[width=0.5\linewidth]{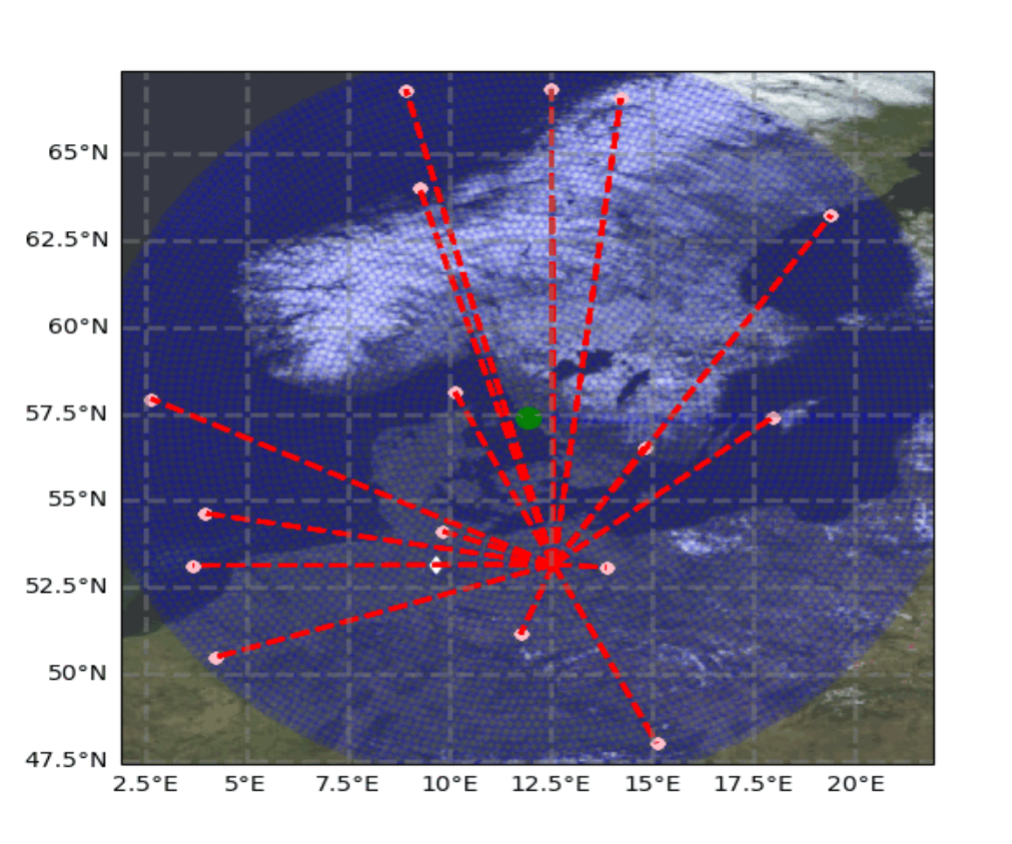}
  \caption{Example of 16 beams (red dotted lines) pointed towards different cells (blue circles) on the ground. The green dot represents the position of the radio telescope.}
  \label{fig:satellite beams}
\end{figure}

Ideally one would like to make a measurement of the radiation characteristics of each satellite in a constellation, so that the total power incident at a radio telescope antenna could be computed. This is very complex and almost impossible due to the facts explained before (dynamic changing of pointing direction of beams) and the fact that nonGSO satellites are rapidly moving across the sky.
Given this impossibility, a simulation model was developed. The model simulates the behaviour of each steerable beam by considering one static beam pattern and mathematically rotating it to point to any possible cell on the ground (see \ref{fig:satellite beams}). The simulation includes randomly sampling the pointing direction, with some constraints as defining the maximum steering angle of a satellite and that a cell can be serviced only by one beam per channel at a time.

The radio telescope beam, or gain, is not considered in this simulation, as the primary parameter of interest is the power flux density incident from each satellite. The gain of the radio telescope antenna can be applied in subsequent stages to calculate the received power based on the satellite's position in its antenna beam pattern.

\begin{figure}[h]
  \center
  \includegraphics[width=0.5\linewidth]{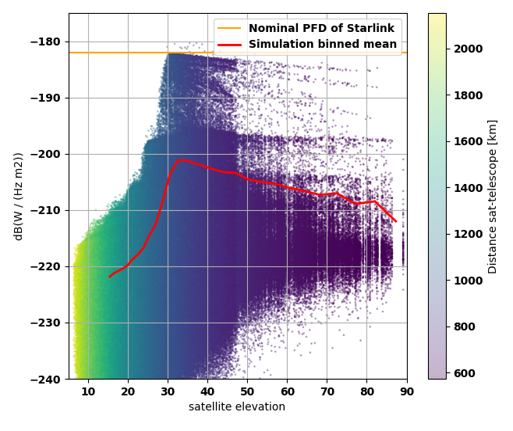}
  \caption{Calculated PFD of individual satellites for each random pointing as a function of satellite elevation.}
  \label{fig:Sim_results}
\end{figure}

The orbital positions of Starlink phase 1 constellation were calculated for 1000 seconds in 5 second time-steps (200 samples) using its orbital parameters and the open-source Python package cysgp4, which is available under GPL-v3 license  \citep{cysgp4}. 1000 random pointing directions (of 16 beams per satellite) were calculated for each visible satellite, and the total power produced by them was added as a linear sum. This resulted in 2e6 power flux density samples per visible satellite.  All samples of  PFD as a function of satellite elevation are presented in \ref{fig:Sim_results}.

\begin{figure}[h]
  \center
  \includegraphics[width=0.5\linewidth]{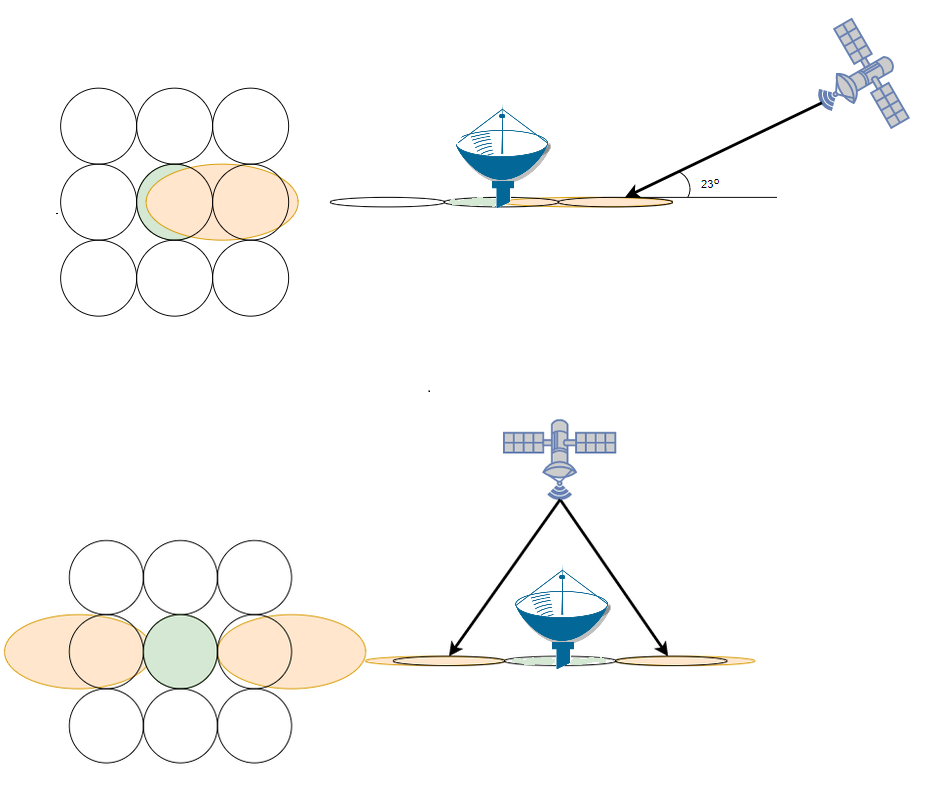}
  \caption{Top view of a rectangular arrangement of cells with the observer in the central cell (left), orange depicts the footprint of a circular beam being elongated due to the slant direction of incidence. Side view of illumination of cells adjacent to the observer from a slant angle (top right) or overhead (bottom right). }
  \label{fig:spillover}
\end{figure}

The simulation result reveal interesting patterns. One is the PFD values reaching a maximum around 30 degrees elevation and then falling at lower elevations. This is consistent with expectations, as satellites at elevations less than about 30 degrees cannot point their beams towards the observer cell or nearby cells due to their maximum steering angle. Another interesting result is that the PFD decreases as the satellites rise in elevation, this is a reasonable result as a satellite high in the sky is more likely to point its beams to cells that are outwards from the radio telescope cell (see Fig~\ref{fig:spillover}). Satellites with low elevations have elongated beams which can spill over into the radio-telescope's cell. This is unavoidable to some degree, even if it is a region masked by the constellation.

\section{Observations}

\begin{figure}[h]
  \center
  \includegraphics[width=1\linewidth]{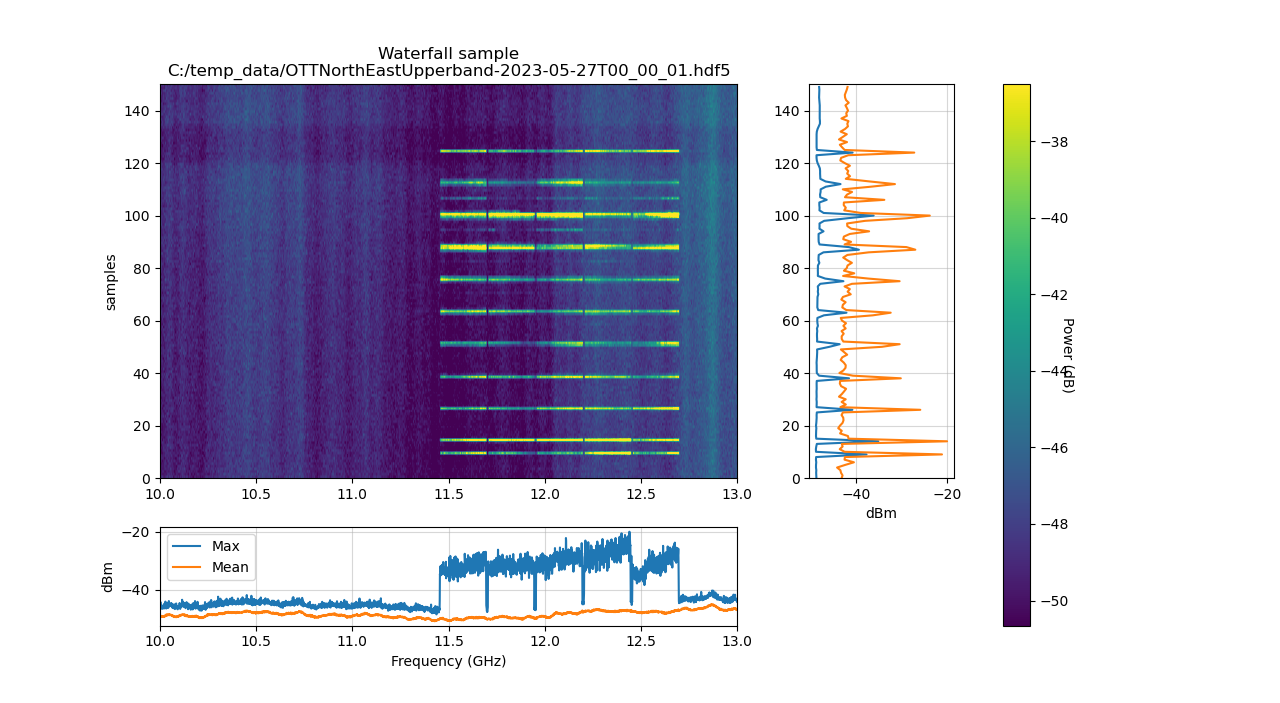}
  \caption{Waterfall plot of received power (top), the high frequency of occurrence of sattelite downlinks in samples (of 1.5~s approximately) is due to the pointing schedule used, to maximise the satellite transits for an observation time. The bottom panel shows the integrated power in time as a function of frequency, clearly showing the 5 channels of 250 MHz used by the satellites. On the right panel is the integrated power in frequency as a function of samples (time).}
  \label{fig:waterfall_measurement}
\end{figure}

Observations with the Onsala Twin Telescopes (OTT) \citep{2019VLBI}, located at the Onsala Space Observatory (OSO) in Sweden, were carried out to validate the simulation model and gain further insights into the impact of satellite constellations on radio astronomy. The OTT consists of two 13.2m paraboloids with receivers that have a noise temperature of 15K from 2.2-14~GHz and 3-18~GHz. A spectrum analyser was connected to the output of each telescope capturing power in a sweep spectral mode with a resolution bandwidth of 1~MHz and a capture time of approximately 1.5~s. Each frequency sweep was transferred from the spectrum analyser to a computer in units of dBm, an example of an observation run can be see in \ref{fig:waterfall_measurement}.

\begin{figure}[h]
  \center
  \includegraphics[width=0.5\linewidth]{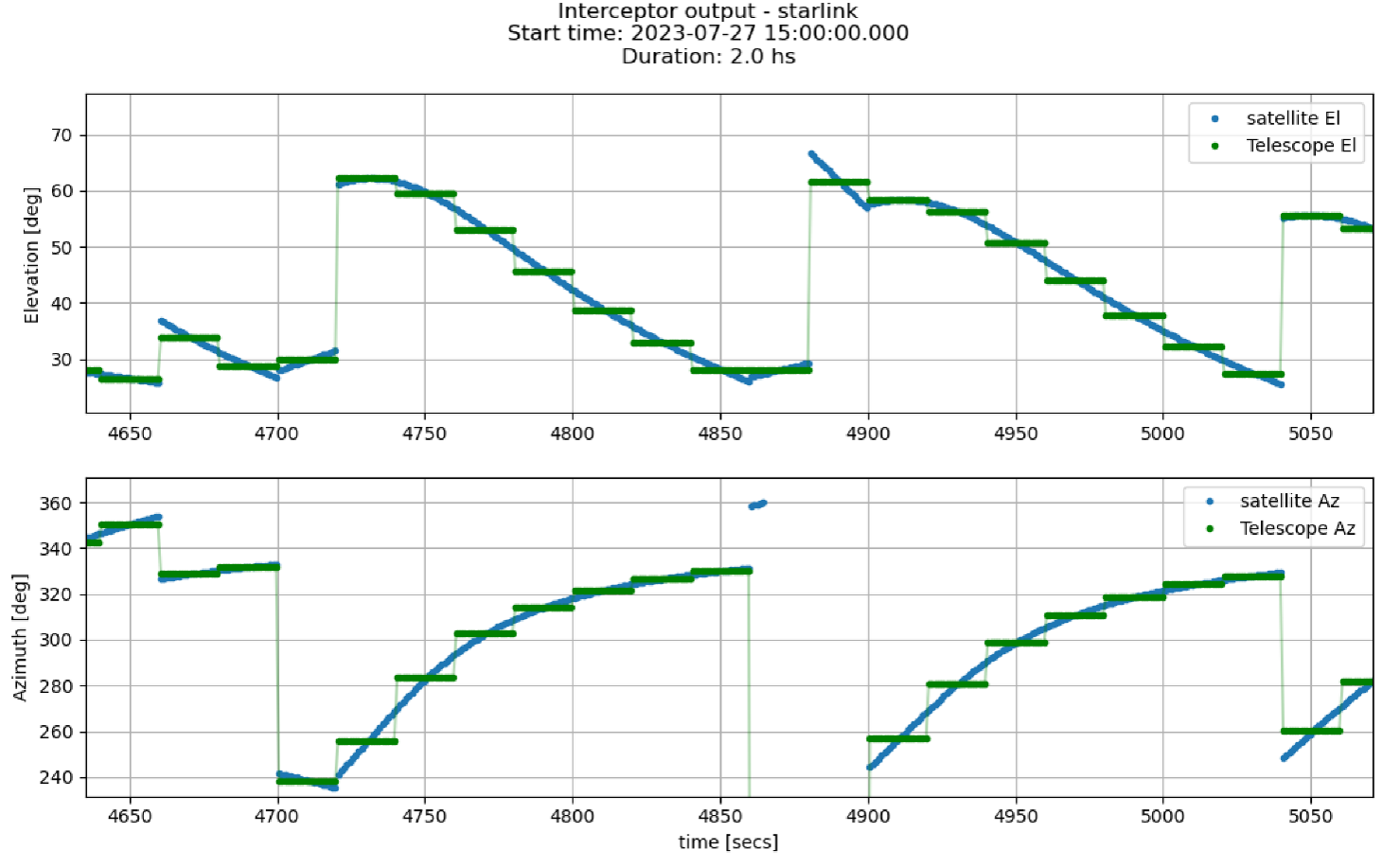}
  \caption{Pointing schedule as a function of time for the Onsala Twin Telescopes in Azimuth (top) and Elevation (bottom). The green dots represent the telescope position and the blue dots present the moving satellites positions.}
  \label{fig:Pointing_plan}
\end{figure}

The telescopes' pointing schedule was calculated using a python script that utilised supplemental Two-Line Element sets (TLEs) calculated by Celestrak and propagated using cysgp4. Both radio telescopes were pointed in directions in the sky which intersect satellite traces, to account for inaccuracies in the satellites' position predictions, a window of +- 20 seconds was used as shown in Fig. \ref{fig:Pointing_plan}. The observations captured over 700 transits of Starlink satellites, where a "transit" is the passing of a satellite through the -3dB beam of the radio telescopes.

One important consideration is that the Starlink system does not illuminate the Onsala Space Observatory, as can be seen in their "Starlink availability map" available at \footnote{\url{https://www.starlink.com/map}} . To account for this the central cell of the simulation was marked as "passivated" in the software so that no beam was pointed towards it.

The power measured by the spectrum analyser is at a point in the signal chain and requires calibration to reference it to the power received by the antenna aperture and the incident power flux density. To calibrate the measured power into PFD, the radio telescopes were pointed towards Cassiopeia A, a source with a known flux density at 11~GHz,  for 2 minutes in every observation run. The calibrated observed results not only provided statistical information on PFD but also revealed the occupancy of different channels by the observed satellites (see Fig~\ref{fig:waterfall_measurement}). Notably, channels 1 and 2 were not used, consistent with the known practice of operators of avoiding these channels to protect adjacent radio astronomy bands \citep{ecc_report_271}. 
\begin{figure}[h]
  \center
  \includegraphics[width=0.5\linewidth]{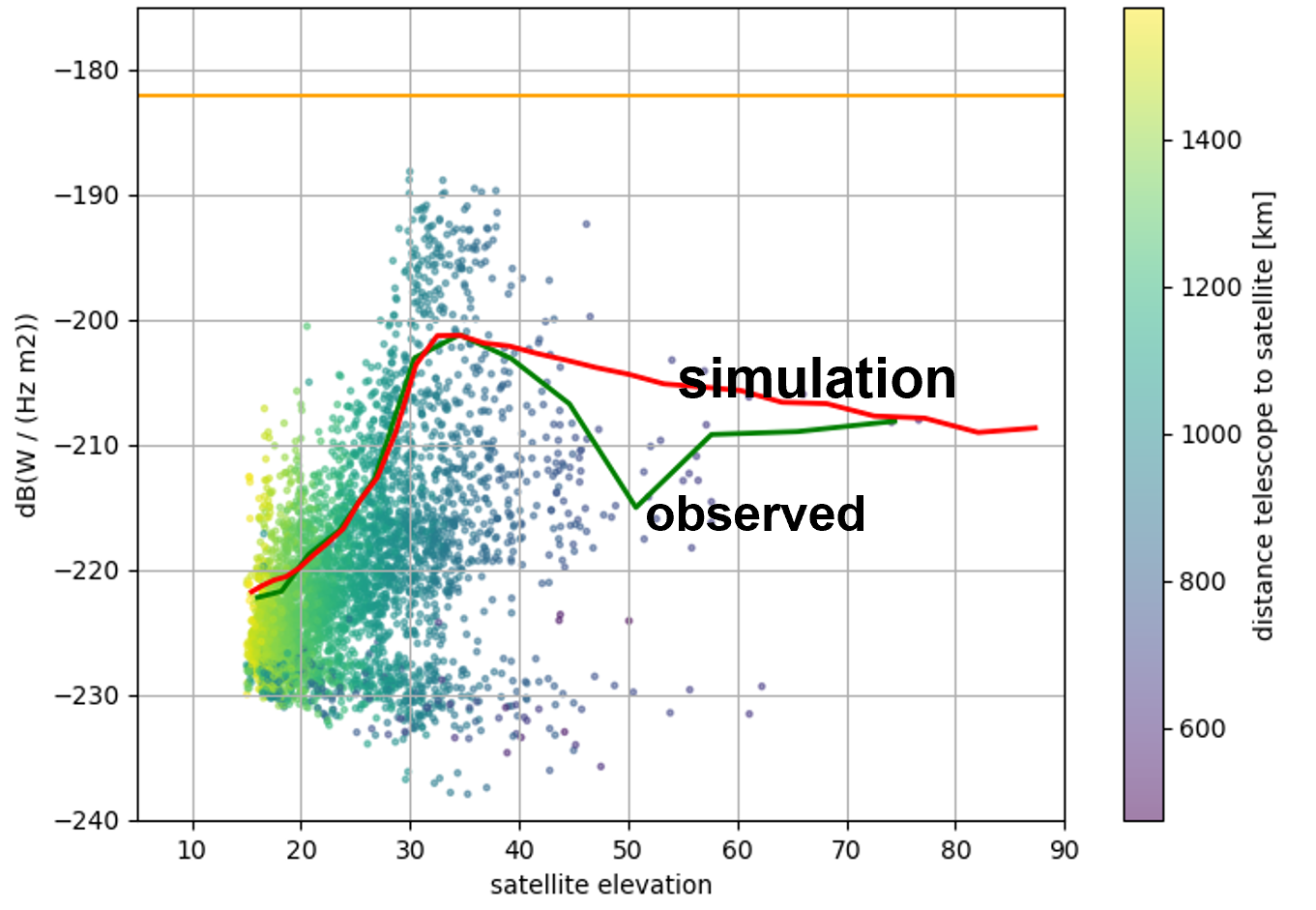}
  \caption{Distribution of the measured PFD for each one of the satellite transits detected. The green curve shows the averaged PFD for  each elevation bin. The red curve is the averaged PFD for each elevation point from the simulation. }
  \label{fig:Measured_results}
\end{figure}

\begin{figure}[h]
  \center
  \includegraphics[width=0.75\linewidth]{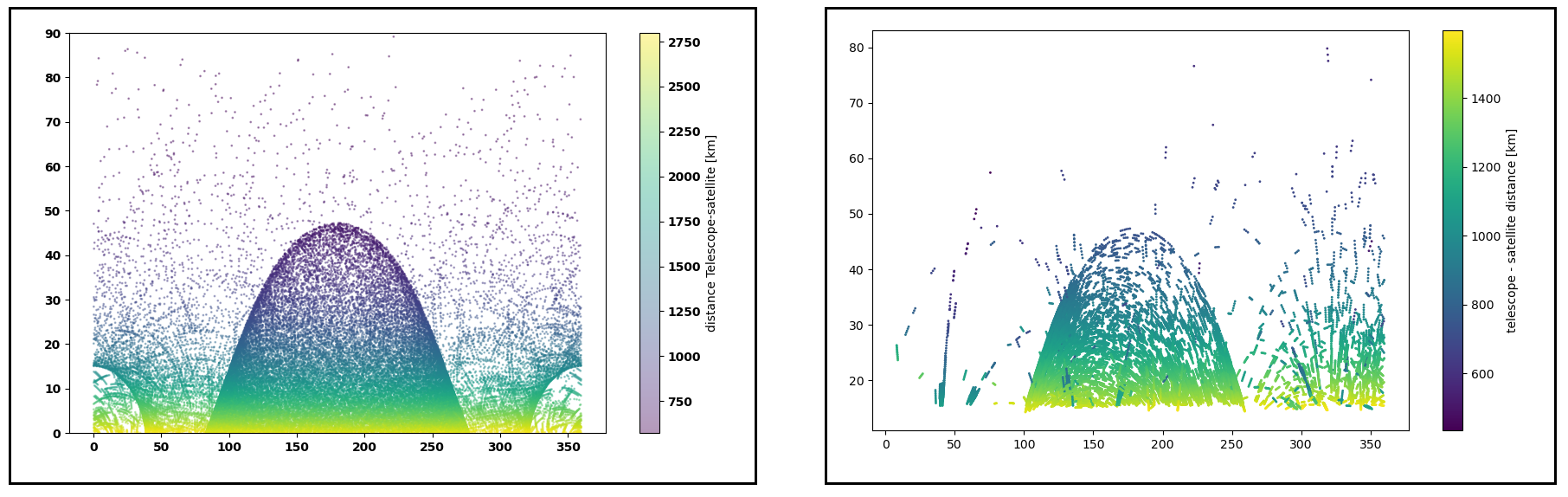}
  \caption{Positions in the sky of the satellites in the simulation (left) and observations with the OTT (right). The range from the observer to the satellite is color coded in km.}
  \label{fig:observed_simulated_satellites}
\end{figure}

The observed mean results in Fig~\ref{fig:Measured_results} closely match the mean result of the simulation for the lower elevations, in higher elevations this correlation does not hold we believe due to a lack of data points in the observations. Due to the latitude of the Onsala telescopes, most satellites are located at low altitude as can be seen in Fig~\ref{fig:observed_simulated_satellites}, to make a substantial increase in observations at high altitudes we would require much more telescope time. The results show that the observations are in good agreement with the model predictions adjusting the parameters. These include  the single beam pattern, number of beams per channel, maximum steering angle for one beam, cell size on the ground, network considerations, and avoidance cells on the ground. The successful alignment of the simulation with observations underscores the model's reliability.

\section{Further applications}
The developed model holds potential for further applications, such as for the simplification of epfd calculations. When the size of the constellations is too large, adding another degree of freedom as the pointing direction of beams could produce the calculation very computation-expensive. This model can also be used to assess the overall occupancy of channels in radio astronomy receivers and their potential impact. By combining the statistical distribution of satellites in the sky with the model results, it becomes possible to estimate the total received power into the radio telescope, either as a full-sky average or as a function of the direction in the sky.

\section*{Acknowledgements}
The authors express their gratitude to the Geodesy Group at the Onsala Space Observatory for providing time on the Twin Telescopes for this experiment, as well as their invaluable assistance in setting up and monitoring observations. Special thanks are extended to Starlink and the Monitoring Group of the Committee on Radio Astronomy Frequencies (CRAF) for constructive discussions during the development of the model. 

\bibliographystyle{iaulike}
\bibliography{refs.bib}

\end{document}